# Light Baryons Mass in a Non-Relativistic Quark Model within an Hypercentral Power Low Potential

NASRIN SALEHI [*]

*Department of Basic Sciences, Shahrood Branch, Islamic Azad University, Shahrood, Iran.*

E-mail: [*]salehi@shahroodut.ac.ir

**Abstract**- In this work, we calculated the baryon mass within a non-relativistically quark model using an approach based on the Gürsey Radicati mass formula (GR). The average energy value of each SU(6) multiplet is described using the SU(6) invariant interaction given by a hypercentral potential. In our series studies we investigate different interactions and situations to gain the best possible model. This goal can be obtained by checking and studying various potentials in different situations. In this paper we present the solution of the Schrödinger equation with an hypercentral power low potential. The results of our model (the combination of our proposed hypercentral Potential and generalized GR mass formula to description of the spectrum) show that the strange and non-strange baryons spectra are in general fairly well reproduced. The overall good description of the mass which we obtain shows that our model can also be used to give a fair description of the energies of the excited multiplets up to three GeV and the position of the Roper resonances of the nucleon.

## 1. Introduction

All the baryons have been made up of three constituent confined quarks. Since the quarks are fermions, the state function for any baryon must be antisymmetric under interchange of any two equal-mass quarks (up and down quarks in the limit of isospin symmetry). The Constituent Quark Models (CQMs) have been recently widely applied to the description of baryon properties [1-5] and most attention has been devoted to the spectrum [6-8]. The baryon spectrum is usually described well, although the various models are quite different. Common to these models is the fact that the three Quark interaction can be divided in two parts: the first one, containing the confinement interaction, is spin and flavour independent and it is therefore $SU(6)$ invariant, while the second violates the $SU(6)$ symmetry [9-11]. One of the most popular ways to violate the $SU(6)$ invariance was the introduction of a hyperfine (spin-spin) interaction [12,13], however in many studies a spin and isospin [1,14,15] or a spin and flavour dependent interaction [1,13] has been considered. It is well known that the Gürsey Radicati mass formula [16] describes quite well the way $SU(6)$ symmetry is broken, at least in the lower part of the baryon spectrum. In this paper we applied the generalized Gürsey Radicati (GR) mass formula which is presented by Giannini and et al [17] to obtain the best description of the strange and nonstrange baryons spectrum. The model we used is a simple CQM where the $SU(6)$ invariant part of the Hamiltonian is the same as in the hypercentral Constituent Quark Model (hCQM) [18, 19] and where the $SU(6)$ symmetry is broken by a generalized GR mass formula.

In sect.2 we remind the hypercentral Constituent Quark Model (hCQM) and then we present the solution of the Schrödinger equation for the octic potential by using the Ansatz method. In the Ansatz approach, which is followed here, we introduce a solution consistent with the requirements of quantum mechanics and thereby the differential equation under study is solved [20, 21]Roughly speaking, we see these analytical tools in all wave equations of quantum mechanics as in many cases the relativistic or nonrelativistic equations appear as Schrödinger-like equation [21, 22]. In the third section in order to describe the splitting within the $SU(6)$ multiplets we introduce the Gürsey Radicati mass formula and generalized GR mass formula in the hCQM, then we give the results obtained by fitting the generalized GR mass formula parameters to the strange and nonstrange baryons energies and we compare the spectrum with the experimental data. Finally, in sect.4 there are some discussions and conclusions.

## 2. Analytical Solution of the Schrödinger Equation for the Octic Potential

We consider baryons as bound states of three quarks. After removing the center of mass coordinate R, the internal quark motion is described by the Jacobi coordinates, $\vec{\rho}$ and $\vec{\lambda}$:

$$\vec{\rho} = \frac{1}{\sqrt{2}}(\vec{r}_1 - \vec{r}_2) \qquad \vec{\lambda} = \frac{1}{\sqrt{6}}(\vec{r}_1 + \vec{r}_2 - 2\vec{r}_3) \qquad (1)$$

or, equivalently, $\rho$, $\Omega_\rho$, $\lambda$, $\Omega_\lambda$. Such that

$$m_\rho = \frac{2m_1 m_2}{m_1 + m_2} \qquad m_\lambda = \frac{3m_3(m_1 + m_2)}{2(m_1 + m_2 + m_3)} \qquad (2)$$

Here $m_1$, $m_2$ and $m_3$ are the constituent quark masses.

In order to describe three - quark dynamics, it is convenient to introduce the hyperspherical coordinates, which are obtained by substituting the absolute values $\rho$ and $\lambda$ by:

$$x = \sqrt{\rho^2 + \lambda^2} \qquad \xi = \arctan(\frac{\rho}{\lambda}) \qquad (3)$$

where $x$ is the hyperradius and $\xi$ the hyperangle.

The search for exact solutions to quantum-mechanical models with rational potentials has been a very significant research aspect in the past decades [23-26]. However it is well recognized that only a very limited number of models in quantum mechanics can be solved exactly. The hypercentral potentials could be of any form. Among them the most frequently studied ones are perhaps the quartic and sextic potentials, both of which allow an sl(2) algebraization [27-29]. Models with higher order anharmonic potentials have applications in the study of structural phase transitions [30], polaron formation in solids [31] and false vacuo in field theory [32]. In our model the interaction potential as the octic potential:

$$V(x) = \sum_{k=1}^{8} a_k x^k = -a_1 x + a_2 x^2 - a_3 x^3 + a_4 x^4 - a_5 x^5 + a_6 x^6 - a_7 x^7 + a_8 x^8 \qquad (4)$$

Therefore, $V(x)$ is in general a three-body potential, since the hyperradius $x$ depends on the coordinates of all three quarks. First we will solve the Schrödinger equation with octic potential by means of the ansatz method, and give the closed-form expressions for the energies then by using the generalized GR mass formula we can try to find the baryons mass. For hypercentral potentials, the Schrödinger equation, in the hyperspherical coordinates, is simply reduced to a single hyperradial equation, while the angular and hyperangular parts of the 3q-states are the known hyperspherical harmonics [33].

Therefore the Hamiltonian will be:

$$H = \frac{P_\rho^2}{2m} + \frac{P_\lambda^2}{2m} + V(x) \qquad (5)$$

and the hyperradial wave function $\psi_{\nu\gamma}(x)$ is determined by the hypercentral Schrödinger equation:

$$\left( \frac{d^2}{dx^2} + \frac{5}{x}\frac{d}{dx} - \frac{\gamma(\gamma+4)}{x^2} \right) \psi_{\nu\gamma}(x) = -2m[E - V(x)]\psi_{\nu\gamma}(x) \qquad (6)$$

where $\gamma$ is the grand angular quantum number and given by $\gamma = 2n + l_\rho + l_\lambda$, $n = 0,1,2,...$; $l_\rho$ and $l_\lambda$ are the angular momenta associated with the $\vec{\rho}$ and $\vec{\lambda}$ variables and $\nu$ denotes the number of nodes of the space three quark wave functions. In equation (6) $m$ is the reduced mass [34] which is defined as:

$$m = \frac{2m_\rho m_\lambda}{m_\rho + m_\lambda} \qquad (7)$$

Now we want to solve the hyperradial Schrödinger equation for the three-body potential interaction (4). The transformation

$$\psi_{\nu\gamma}(x) = x^{-\frac{5}{2}} \varphi_{\nu\gamma}(x) \qquad (8)$$

reduces Eq. (6) to the form:

$$\varphi''_{\nu\gamma}(x) + \left[ \varepsilon + b_1 x - b_2 x^2 + b_3 x^3 - b_4 x^4 + b_5 x^5 - b_6 x^6 + b_7 x^7 - b_8 x^8 - \frac{(2\gamma+3)(2\gamma+5)}{4x^2} \right] \varphi_{\nu\gamma}(x) = 0 \qquad (9)$$

The hyperradial wave function $\varphi_{\nu\gamma}(x)$ is a solution of the reduced Schrödinger equation for each of the three identical particles with the mass $m$ and interacting potential (4), where

$$\varepsilon = 2mE, b_1 = 2ma_1, b_2 = 2ma_2, b_3 = 2ma_3, b_4 = 2ma_4, b_5 = 2ma_5, b_6 = 2ma_6, b_7 = 2ma_7, b_8 = 2ma_8$$
(10)

We suppose the following form for the wave function:
$$\varphi_{\nu\gamma} = h(x) e^{g(x)} \tag{11}$$

Now for the functions $h(x)$ and $g(x)$ we make use of the ansatz [35, 36]:
$$h(x) = \prod_{i=1}^{\nu}(x - \alpha_i^{\nu}) \qquad \nu = 1, 2, ...$$
$$h(x) = 1 \qquad \nu = 0 \tag{12}$$

$$g(x) = -\frac{1}{2}\alpha x^2 + \frac{1}{3}\beta x^3 - \frac{1}{4}\tau x^4 + \frac{1}{5}\delta x^5 + k\ln x$$

By calculating $\varphi_{\nu\gamma}''(x)$ from Eq. (11) and comparing with Eq. (9), we obtain:

$$\left[-b_1 x + b_2 x^2 - b_3 x^3 + b_4 x^4 - b_5 x^5 + b_6 x^6 - b_7 x^7 + b_8 x^8 + \frac{(2\gamma+3)(2\gamma+5)}{4x^2} - \varepsilon\right] =$$
$$\frac{h''(x) + 2g'(x)h'(x)}{h(x)} + g''(x) + g'^2(x) \tag{13}$$

By substituting Eq. (12) in to Eq. (13) we obtain the following equation for $\nu = 0$

$$-b_1 x + b_2 x^2 - b_3 x^3 + b_4 x^4 - b_5 x^5 + b_6 x^6 - b_7 x^7 + b_8 x^8 + \frac{(2\gamma+3)(2\gamma+5)}{4x^2} - \varepsilon =$$
$$\delta^2 x^8 - 2\tau\delta x^7 + (2\beta\delta + \tau^2)x^6 - (2\beta\tau + 2\alpha\delta)x^5 + (2\alpha\tau + \beta^2)x^4 - (2\alpha\beta - 4\delta - 2\delta k)x^3 + \tag{14}$$
$$(\alpha^2 - 3\tau - 2\tau k)x^2 - (-2\beta - 2\beta k)x - 2\alpha k - \alpha + \frac{1}{x^2}(k^2 - k)$$

By equating the corresponding powers of $x$ on both sides of Eq. (14), we can obtain

$$-\varepsilon = -2\alpha k - \alpha$$
$$-b_1 = 2\beta + 2\beta k$$
$$b_2 = \alpha^2 - 3\tau - 2k\tau$$
$$-b_3 = -2\alpha\beta + 4\alpha + 2\delta k$$
$$b_4 = 2\alpha\tau + \beta^2$$
$$-b_5 = -2\beta\tau - 2\alpha\delta \tag{15}$$
$$b_6 = 2\beta\delta + \tau^2$$
$$-b_7 = -2\tau\delta$$
$$b_8 = \delta^2$$
$$\frac{(2\gamma+3)(2\gamma+5)}{4} = k^2 - k$$

We find constraints on potential parameters as

$$\delta = \sqrt{b_8} \;,\; \beta = \frac{4b_6 b_8 - b_7^2}{8 b_8 \sqrt{b_8}} \;,\; \tau = \frac{b_7}{2\sqrt{b_8}} \;,\; \alpha = \frac{-4 b_6 b_7 b_8 + b_7^3 + 8 b_5 b_8^2}{16 b_8^2 \sqrt{b_8}} \;,\; \varepsilon = \alpha(1+2k) \;,\; k = \gamma + \frac{5}{2} \quad (16)$$

We have taken the parameters as Eq. (16) to satisfy the quantum behavior of the problem. The energy eigenvalues for the mode $v = 0$ and grand angular momentum $\gamma$ from Eqs. (10) and (16) are given as

$$E_{0\gamma} = \frac{2\sqrt{2m}(-4 a_6 a_7 a_8 + a_7^3 + 8 a_5 a_8^2)(\gamma + 3)}{16 a_8^2 \sqrt{a_8}} \quad (17)$$

We can write the obtained energy (17) versus parameters $b_1, b_2, b_3$ and $b_4$ which are combinations of parameters $b_5, b_6, b_7$ and $b_8$. Therefore, eigenvalue depend on all the 8 mentioned parameters. We obtain the parameters $b_1, b_2, b_3$ and $b_4$ as

$$b_1 = \frac{2 b_7^2 - 4 b_6 b_8 + \left(b_7^2 - 4 b_6 b_8\right)(2\gamma + 5)}{8 b_8 \sqrt{b_8}}$$

$$b_2 = \frac{\left(-4 b_6 b_7 b_8 + b_7^3 + 8 b_5 b_8^2\right)^2 - \left(3 b_7 + (2\gamma + 5) b_7\right)\left(256 b_8^4 \sqrt{b_8}\right)}{\left(16 b_8^2 \sqrt{b_8}\right)^2} \quad (18)$$

$$b_3 = \frac{\left(b_7^2 - 4 b_6 b_8\right)\left(-4 b_6 b_7 b_8 + b_7^3 + 8 b_5 b_8^2\right) - 32\left(4 b_6 b_7 b_8 - b_7^3 - 8 b_5 b_8^2\right) b_8 \sqrt{b_8} + 128(2\gamma + 5) b_8^3 \sqrt{b_8}}{128 b_8^4}$$

$$b_4 = \frac{-16 b_6 b_7^2 b_8 + 4 b_7^4 + 32 b_5 b_7 b_8^2 + \left(4 b_6 b_8 - b_7^2\right)^2}{32 b_8^3}$$

which are constraint equations.

Now let us consider mode $v = 1$. Substituting Eq. (12) into Eq. (13) with $v = 1$ we arrive at

$$-b_1 x^2 + b_2 x^3 - b_3 x^4 + b_4 x^5 - b_5 x^6 + b_6 x^7 - b_7 x^8 + b_8 x^9 + \frac{(2\gamma+3)(2\gamma+5)}{4x} - \varepsilon x +$$

$$\alpha_1^1 b_1 x - \alpha_1^1 b_2 x^2 + \alpha_1^1 b_3 x^3 - \alpha_1^1 b_4 x^4 + \alpha_1^1 b_5 x^5 - \alpha_1^1 b_6 x^6 + \alpha_1^1 b_7 x^7 - \alpha_1^1 b_8 x^8 - \frac{\alpha_1^1 (2\gamma+3)(2\gamma+5)}{4 x^2} + \alpha_1^1 \varepsilon =$$

$$\delta^2 x^9 - 2\tau\delta x^8 + (2\beta\delta + \tau^2) x^7 - (2\beta\tau + 2\alpha\delta) x^6 + (2\alpha\tau + \beta^2) x^5 - (2\alpha\beta - 4\delta - 2\delta k) x^4 +$$
$$-\alpha_1^1 \delta^2 x^8 + 2\alpha_1^1 \tau\delta x^7 - \alpha_1^1 (2\beta\delta + \tau^2) x^6 + \alpha_1^1 (2\beta\tau + 2\alpha\delta) x^5 - \alpha_1^1 (2\alpha\tau + \beta^2) x^4 + \alpha_1^1 (2\alpha\beta - 4\delta - 2\delta k) x^3 +$$
$$\delta^2 x^8 - 2\tau\delta x^7 + (2\beta\delta + \tau^2) x^6 - (2\beta\tau + 2\alpha\delta) x^5 + (2\alpha\tau + \beta^2) x^4 - (2\alpha\beta - 4\delta - 2\delta k) x^3 +$$
$$(\alpha^2 - 3\tau - 2\tau k) x^3 - (-2\beta - 2\beta k) x^2 - 2\alpha k x - \alpha x + \frac{1}{x}(k^2 - k) - \alpha_1^1 (\alpha^2 - 3\tau - 2\tau k) x^2 + \alpha_1^1 (-2\beta - 2\beta k) x$$
$$+ 2\alpha_1^1 \alpha k + \alpha \alpha_1^1 - \frac{1}{x^2} \alpha_1^1 (k^2 - k) - 2\alpha x + 2\beta x^2 - 2\tau x^3 + 2\delta x^4 + \frac{2k}{x} \quad (19)$$

We also obtain the following equations by equating the corresponding powers of $x$ on both sides of Eq. (19)

$$-b_7 - \alpha_1^1 b_8 = -2\tau\delta - \alpha_1^1 \delta^2$$
$$b_6 + \alpha_1^1 b_7 = 2\beta\delta + \tau^2 + 2\alpha_1^1 \tau\delta$$
$$-b_5 - \alpha_1^1 b_6 = -2\beta\tau - 2\alpha\delta - 2\alpha_1^1 \beta\delta - \alpha_1^1 \tau^2$$
$$b_4 + \alpha_1^1 b_5 = 2\alpha\tau + \beta^2 + 2\alpha_1^1 \beta\tau + 2\alpha_1^1 \alpha\delta$$
$$-b_3 - \alpha_1^1 b_4 = -2\alpha\beta + 4\alpha + 2\delta k - 2\alpha\alpha_1^1 \tau - \alpha_1^1 \beta^2 + 2\delta$$
$$b_2 + \alpha_1^1 b_3 = \alpha^2 - 3\tau - 2k\tau + 2\alpha\alpha_1^1 \beta - 4\alpha\alpha_1^1 - 2\alpha_1^1 \delta k - 2\tau \tag{20}$$
$$-b_1 - \alpha_1^1 b_2 = 2\beta + 2\beta k - \alpha_1^1 \alpha^2 + 3\alpha_1^1 \tau + 2\alpha_1^1 \tau k + 2\beta$$
$$-\varepsilon + \alpha_1^1 b_3 = -2\alpha k - \alpha - 2\alpha_1^1 \beta - 2\alpha_1^1 \beta k - 2\alpha$$
$$\alpha_1^1 \varepsilon = 2\alpha\alpha_1^1 k + \alpha\alpha_1^1$$
$$\frac{(2\gamma+3)(2\gamma+5)}{4} = k^2 + k$$

Therefore, we find restrictions on potential parameters as

$$k = \gamma + \frac{5}{2} \quad , \quad \delta = \frac{\sqrt{-b_7 - \alpha_1^1 b_8}}{\sqrt{\alpha_1^1}} \quad , \quad \beta = (\alpha_1^1)^{\frac{3}{2}}\sqrt{-b_7 - \alpha_1^1 b_8} + \sqrt{b_4 + \alpha_1^1 b_5 - (\alpha_1^1)^3 b_7 - (\alpha_1^1)^4 b_8}$$

$$\alpha = \frac{\varepsilon}{2k+1} \tag{21}$$

In the case of $\nu = 1$ the energy eigenvalues are given as

$$E = \frac{\alpha_1^1 2m a_3 + 2\alpha k + 3\alpha + 2\alpha_1^1 \beta + 2\alpha_1^1 \beta k}{2m} \tag{22}$$

using of Eqs. (10) and (21) in terms of grand angular momentum $\gamma$.

## 3. Baryons Mass Splitting by Using the Generalizing Gürsey Radicati Mass Formula

The description of the baryons spectrum obtained by the hypercentral Constituent Quark Model (hCQM)[12] is fairly good and comparable to the results of other approaches, but in some cases the splitting within the various $SU(6)$ multiplets are too low and not all adequately described by the hyperfine interaction. This is particularly true for the Roper resonances. The preceding results [15, 18, 37] shows that both spin and isospin dependent terms in the quark Hamiltonian are important. Description of the splitting within the $SU(6)$ baryon multiplets is provided by the Gürsey Radicati mass formula [16]:

$$M = M_0 + CC_2[SU_S(2)] + DC_1[U_Y(1)] + E[C_2[SU_I(2)] - \frac{1}{4}(C_1[U_Y(1)])^2] \tag{23}$$

where $M_0$ is the average energy value of the $SU(6)$ multiplet, $C_2[SU_S(2)]$ and $C_2[SU_I(2)]$ are the $SU(2)$ Casimir operators for spin and isospin, respectively, and $C_1[U_Y(1)]$ is the Casimir operator for the $U(1)$ subgroup generated by the hypercharge $Y$ [38, 39].
This mass formula has tested to be successful in the description of the ground state baryon masses, however, as stated by the authors themselves, it is not the most general mass formula that can be written on the basis of a broken $SU(6)$ symmetry. In order to generalize Eq. (23), Giannini and et al considered a dynamical spin-flavor symmetry $SU_{SF}(6)$ [17] and described the $SU_{SF}(6)$ symmetry breaking mechanism as:

$$M = M_0 + A C_2[SU_{SF}(6)] + B C_2[SU_F(3)] + C C_2[SU_S(2)] + D C_1[U_Y(1)] \qquad (24)$$
$$+ E[C_2[SU_I(2)] - \frac{1}{4}(C_1[U_Y(1)])^2]$$

In Eq. (24) the spin term represents spin-spin interactions, the flavor term denotes the flavor dependence of the interactions, and the $SU_{SF}(6)$ term depends on the permutation symmetry of the wave functions, represents "signature - dependent" interactions. The signature – dependent (or exchange) interactions were extensively investigated years ago within the framework of Regge theory [38]. The last two terms represent the isospin and hypercharge dependence of the masses. The generalized Gürsey Radicati mass formula Eq. (24) can be used to describe the baryons spectrum, provided that two conditions are fulfilled. The first condition is the feasibility of using the same splitting coefficients for different $SU(6)$ multiplets. This seems actually to be the case, as shown by the algebraic approach to the baryon spectrum [1], where a formula similar to Eq. (24) has been applied. The second condition is given by the feasibility of getting reliable values for the unperturbed mass values $M_0$ [17]. For this goal we regarded the $SU(6)$ invariant part of the hCQM, which provides a good description of the baryons spectrum and used the Gürsey Radicati inspired $SU(6)$ breaking interaction to describe the splitting within each $SU(6)$ multiplet.

The baryons masses can be obtain by using of of $H_{GR}$ equation as follows

$$M = 3m + E_{\upsilon\gamma} + A \langle C_2[SU_{SF}(6)] \rangle + B \langle C_2[SU_F(3)] \rangle + C \langle C_2[SU_S(2)] \rangle + D \langle C_1[U_Y(1)] \rangle \qquad (25)$$
$$+ E \left[ \langle C_2[SU_I(2)] \rangle - \frac{1}{4} \langle (C_1[U_Y(1)]) \rangle^2 \right].$$

In order to simplify the solving procedure, the constituent quarks masses are assumed to be the same for up, down and strange quark flavors $(m_u = m_d = m_s = 293 \text{ MeV})$, therefore, within this approximation, the $SU(6)$ symmetry is only broken dynamically by the spin and flavour dependent terms in the Hamiltonian. In previous section we determined eigenenergies $E_{\upsilon\gamma}$ by ansatz solution of the Schrödinger equation for the hypercentral Potential (4). In Figs. (1) and (2), we have investigated the dependence of the ground state energy level on the potential parameters ($a_6, a_7$ and $a_8$) for different values of $a_5$ and the dependence of the ground state energy level on the potential parameters ($a_5, a_6$ and $a_7$) for different values of $a_8$, respectively. For calculating the baryons mass according to Eq. (25), we need to find the unknown parameters. For this purpose we choose a limited number of well-known resonances and express their mass differences using $H_{GR}$ and the Casimir operator expectation values

$$N(1650)S11 - N(1535)S11 = 3C$$
$$4N(938)P11 - \Sigma(1193)P11 - 3\Lambda(1116)P01 = 4D \qquad (26)$$
$$\Sigma(1193)P11 - \Lambda(1116)P01 = 2E.$$

Leading to the numerical values: $C = 38.3$, $D = -197.3$ MeV and $E = 38.5$ MeV. We determined $m$, $a_5$, $a_6$, $a_7$ and $a_8$ ( in Eq. (17) ) and the two coefficients $A$ and $B$ of Eq. (25) in a simultaneous fit to the 3 and 4 star resonances of Table 2 which have been assigned as octet and decuplet states. We have calculated the spectrum of Roper $N(1440)$ and the $\Delta$ (1600) in the case $\nu = 1$ and $\gamma = 0$ (from Eq. (22)). The fitted parameters are reported in Table 1, while the resulting spectrums are shown in Figs. (3) and (4). The corresponding numerical values are given in Table 2, column $M_{ourCalc}$. The percentage of relative error for our calculations is between 0 and 12 % (column 6, in table 2). Comparison between our results and the experimental masses [40] show that the baryon spectrums are, in general, fairly well reproduced.

## 4. Conclusion

In this work we have investigated the mass spectrum of baryons resonances on the non-relativistic limit and we have shown that the generalized Gürsey Radicati mass formula is a good parametrization of the baryon energy splittings coming from $SU(6)$ breaking. In our model, the energy splittings within the $SU(6)$ multiplets are considered as perturbations added to the $SU(6)$ invariant levels, which are given by our suggested hypercentral potential. For reproducing the spectrum of baryons resonances, we calculated the energy eigenvalues by solution of the Schrödinger equation for confining potential. Then, we fitted the generalized GR mass formula parameters to the baryons energies and calculated the baryons mass according to Eq. (24). The overall good description of the spectrum which we obtain shows that our model can also be used to give a fair description of the energies of the excited multiplets with more than 2 GeV mass and not only for the ground state. Moreover, our model reproduces the position of the negative-parity resonance. There are still problems in the reproduction of the experimental masses in $\Delta$ (1620) S31 and $\Sigma$(1670) D13 turn out to have predicted mass about 100 MeV above the experimental value. A better agreement may be obtained either using the square of the mass [1] or trying to include a spatial dependence in the $SU(6)$ breaking part, which may have, among others, a delta or Gaussian factor, in order to decrease the breaking with the increase of the spatial excitation[17].

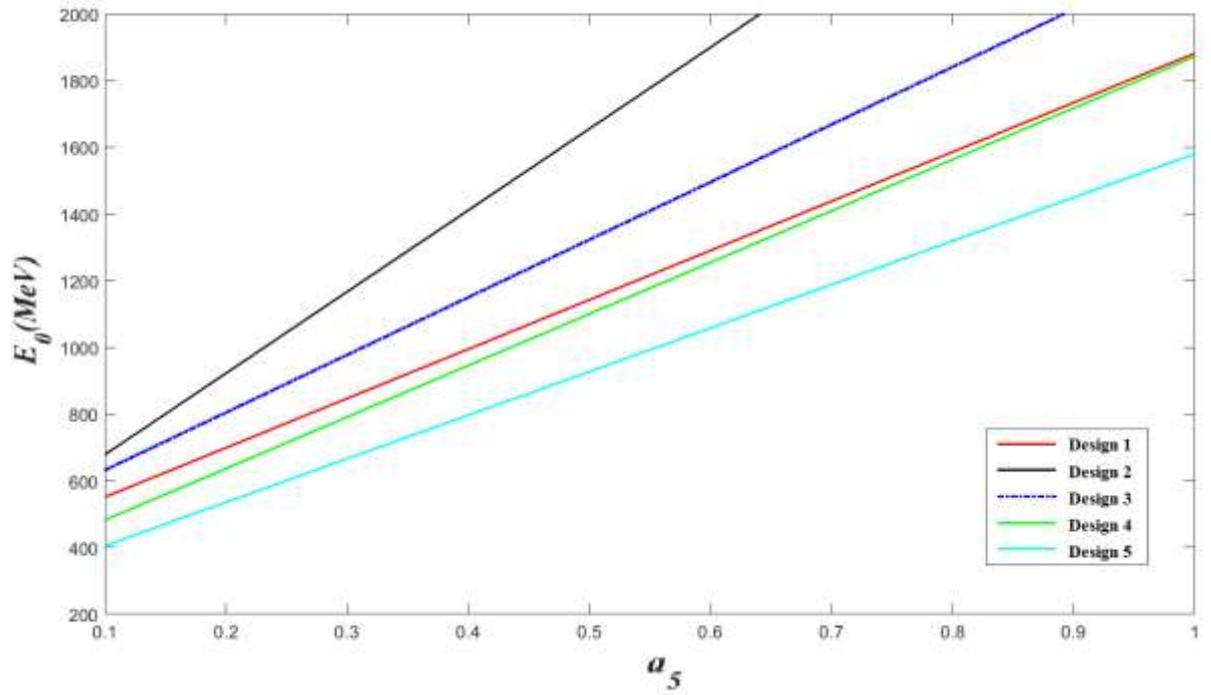

**Fig.1.** The dependence of the ground state energy level (Eq. (17)) on the potential parameters ($a_6, a_7$ and $a_8$) for different values of $a_5$ [ **Design 1**: $a_6 = 0.481, a_7 = 0.381, a_8 = 0.546$. **Design 2**: $a_6 = a_7 = a_8 = 0.2$. **Design 3**: $a_6 = 0.2, a_7 = 0.3, a_8 = 0.4$. **Design 4**: $a_6 = a_7 = a_8 = 0.5$. **Design 5**: $a_6 = a_7 = a_8 = 0.7$ ].

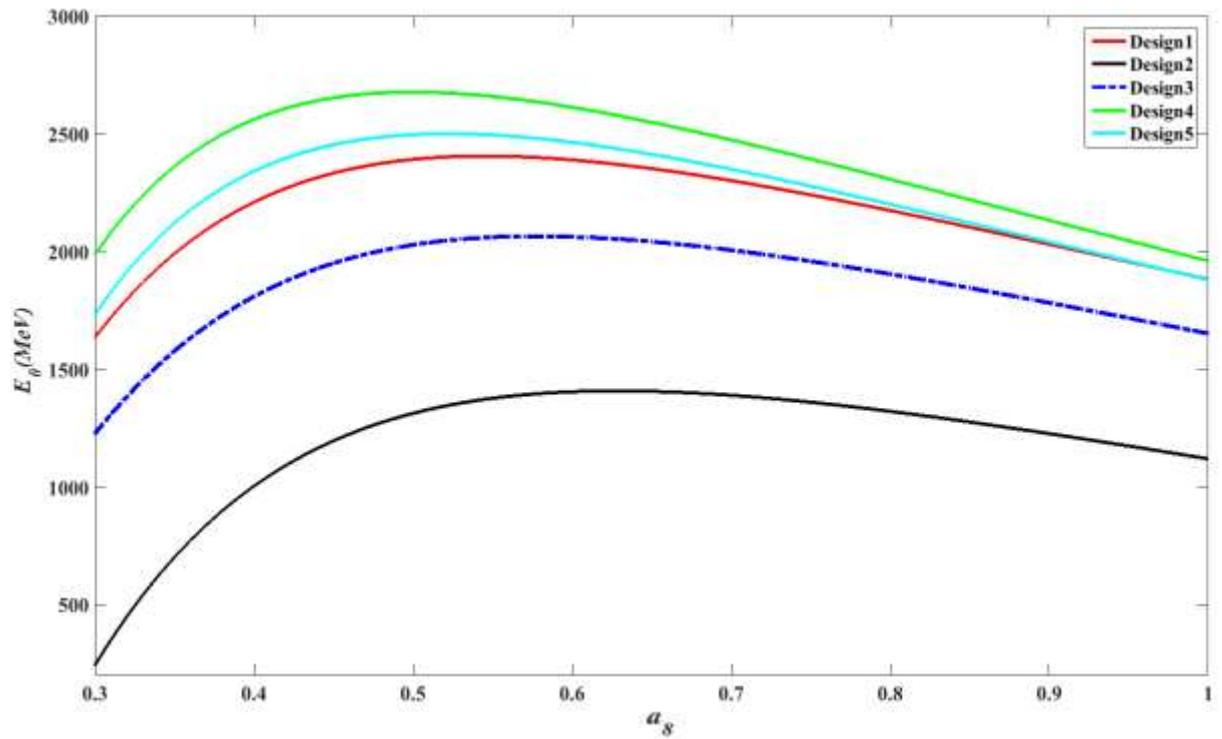

**Fig. 2.** The dependence of the ground state energy level (Eq. (17)) on the potential parameters ($a_5, a_6$ and $a_7$) for different values of $a_8$ [ **Design 1**: $a_5 = 0.388, a_6 = 0.481, a_7 = 0.381$. **Design 2**: $a_5 = 0.35, a_6 = 0.481$,

$a_7 = 0.381$. **Design 3**: $a_5 = 0.38, a_6 = 0.481, a_7 = 0.4$. **Design 4**: $a_5 = 0.38, a_6 = 0.481, a_7 = 0.32$. **Design 5**: $a_5 = 0.38, a_6 = 0.48, a_7 = 0.34$ ].

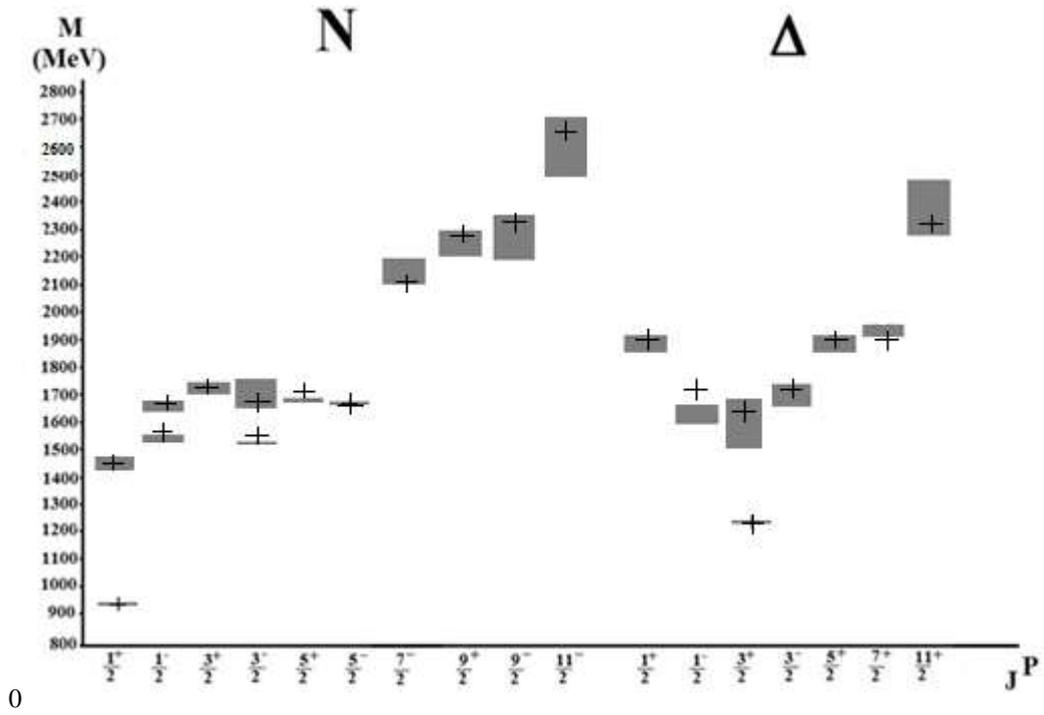

**Fig. 3.** Comparison between the experimental mass spectrum of three and four star N and Δ resonances [40] (gray boxes) and our calculated masses (+) which obtained with the Eq. (25) fixing the mass relation parameters by a fitting procedure.

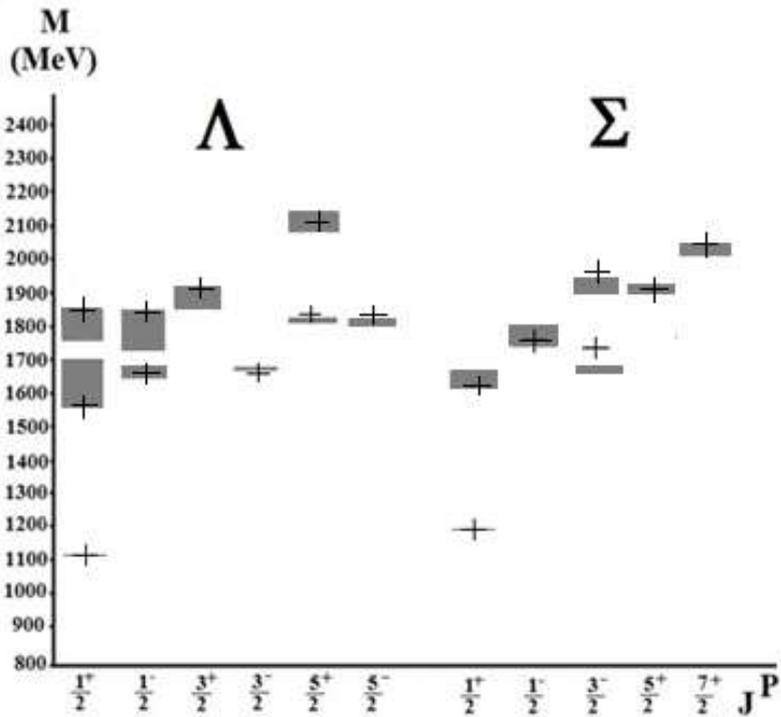

**Fig. 4.** Comparison between the experimental mass spectrum of three and four star Λ, Σ and Ξ resonances [40] (gray boxes) and our calculated masses (+) which obtained with the Eq. (25) fixing the mass relation parameters by a fitting procedure.

Table 1. The fitted values of the parameters of the Eq. (25) for N, Δ, Λ and Σ baryons, obtained with resonances mass differences and global fit to the experimental resonance masses [40].

| Parameter | A | B | C | D | E | m | $a_5$ | $a_6$ | $a_7$ | $a_8$ |
|---|---|---|---|---|---|---|---|---|---|---|
| Value | -16.735 MeV | 20.012 MeV | 38.3 | -197.3 MeV | 38.5 MeV | 293 MeV | 0.388 | 0.481 | 0.381 | 0.546 |

Table 2. Mass spectrum of baryons resonances (in MeV) calculated with the mass formula Eq. (25). The column $M_{Our\ Calc}$ contains our calculations with the parameters of table 1 and column 6 indicate the percentage of relative error for our calculations.

| Baryon | Status | Mass(exp)[40] | State | $M_{Our\ Calc}$ | Percent of relative error |
|---|---|---|---|---|---|
| N(938) P11 | **** | 938 | $^2 8_{1/2}[56, 0^+]$ | 938 | 0% |
| N(1440) P11 | **** | 1410-1450 | $^2 8_{1/2}[56, 0^+]$ | 1423.21 | 0.93% - 1.84% |
| N(1520) D13 | **** | 1510-1520 | $^2 8_{3/2}[70, 1^-]$ | 1549.15 | 2.59% - 1.91% |
| N(1535) S11 | **** | 1525-1545 | $^2 8_{1/2}[70, 1^-]$ | 1549.15 | 1.58% - 0.26% |
| N(1650) S11 | **** | 1645-1670 | $^4 8_{1/2}[70, 1^-]$ | 1664.05 | 1.15% - 0.35% |
| N(1675) D15 | **** | 1670-1680 | $^4 8_{5/2}[70, 1^-]$ | 1664.05 | 0.35% - 0.94% |
| N(1680) F15 | *** | 1680-1690 | $^2 8_{5/2}[56, 2^+]$ | 1723.33 | 2.57% - 1.97% |
| N(1700) D13 | *** | 1650-1750 | $^4 8_{3/2}[70, 1^-]$ | 1664.05 | 0.85% - 4.91% |
| N(1720) P13 | **** | 1700-1750 | $^2 8_{3/2}[56, 2^+]$ | 1723.33 | 1.37% - 1.52% |
| N(2190) G17 | **** | 2100-2200 | $^2 8_{7/2}[70, 3^-]$ | 2110.1 | 0.48% - 4.08% |
| N(2220) H19 | **** | 2200-2300 | $^2 8_{9/2}[56, 4^+]$ | 2284.27 | 3.8% - 0.68% |
| N(2250) G19 | **** | 2200-2350 | $^4 8_{9/2}[70, 3^-]$ | 2337.19 | 6.23% - 0.54% |
| N(2600) I1,11 | *** | 2550-2750 | $^2 8_{11/2}[70, 5^-]$ | 2671.04 | 4.74% - 2.87% |
| Δ (1232) P33 | **** | 1230-1234 | $^4 10_{3/2}[56, 0^+]$ | 1228.44 | 0.12% - 0.45% |
| Δ (1600) P33 | *** | 1500-1700 | $^4 10_{3/2}[56, 0^+]$ | 1642.51 | 9.5% - 3.38% |
| Δ (1620) S31 | **** | 1600-1660 | $^2 10_{1/2}[70, 1^-]$ | 1724.69 | 7.79% - 3.89% |
| Δ (1700) D33 | **** | 1670-1750 | $^2 10_{3/2}[70, 1^-]$ | 1724.69 | 3.27% - 1.44% |
| Δ (1905) F35 | **** | 1855-1910 | $^4 10_{5/2}[56, 2^+]$ | 1901.57 | 2.51% - 0.44% |
| Δ (1910) P31 | **** | 1860-1910 | $^4 10_{1/2}[56, 2^+]$ | 1901.57 | 2.23% - 0.44% |
| Δ (1950) F37 | **** | 1915-1950 | $^4 10_{7/2}[56, 2^+]$ | 1901.57 | 0.7% - 2.48% |
| Δ (2420) H3, 11 | **** | 2300-2500 | $^4 10_{11/2}[56, 4^+]$ | 2350.33 | 2.18% - 5.98% |
| Λ (1116)P01 | **** | 1116 | $^2 8_{1/2}[56, 0^+]$ | 1116.05 | 0.004% |
| Λ (1600)P01 | *** | 1560-1700 | $^2 8_{1/2}[56, 0^+]$ | 1564.81 | 0.3% - 7.99% |
| Λ (1670)S01 | **** | 1660-1680 | $^2 8_{1/2}[70, 1^-]$ | 1671.11 | 0.66% - 0.52% |
| Λ (1690)D03 | **** | 1685-1695 | $^2 8_{3/2}[70, 1^-]$ | 1671.11 | 0.82% - 1.4% |
| Λ (1800)S01 | *** | 1720-1850 | $^4 8_{1/2}[70, 1^-]$ | 1842.1 | 7.09% - 0.42% |
| Λ (1810)P01 | *** | 1750-1850 | $^2 8_{1/2}[70, 0^+]$ | 1839.39 | 5.1% - 0.57% |
| Λ (1830)D05 | **** | 1810-1830 | $^4 8_{5/2}[70, 1^-]$ | 1842.1 | 1.77% - 0.66% |
| Λ (1890)P03 | **** | 1850-1910 | $^2 8_{3/2}[56, 2^+]$ | 1901.38 | 2.77% - 0.45% |
| Λ (2110)F05 | **** | 2090-2140 | $^4 8_{5/2}[70, 2^+]$ | 2122.57 | 1.55% - 0.81% |
| Σ (1193) P11 | **** | 1193 | $^2 8_{1/2}[56, 0^+]$ | 1193.05 | 0.004% |
| Σ (1660)P11 | *** | 1630-1690 | $^2 8_{1/2}[56, 0^+]$ | 1641.81 | 0.72% - 2.85% |
| Σ (1670)D13 | **** | 1665-1685 | $^2 8_{3/2}[70, 1^-]$ | 1748.11 | 4.99% - 3.74% |
| Σ (1750)S11 | *** | 1730-1800 | $^2 8_{1/2}[70, 1^-]$ | 1748.11 | 1.04% - 2.88% |
| Σ (1915)F15 | **** | 1900-1935 | $^2 8_{5/2}[56, 2^+]$ | 1922.28 | 1.17% - 0.65% |
| Σ (1940)D13 | *** | 1900-1950 | $^2 8_{3/2}[56, 1^-]$ | 1978.38 | 4.12% - 1.45% |
| Σ *(2030)F17 | **** | 2025-2040 | $^4 10_{7/2}[56, 2^+]$ | 2041.12 | 0.79% - 0.05% |